\documentclass[aip,apl,twocolumn,reprint]{revtex4-1}
\usepackage{graphicx}

\begin{document}

\title{Modulation and amplification of radiative far field heat transfer : towards a simple radiative thermal transistor}
\author{Karl Joulain,Youn\`es Ezzahri, J\'er\'emie Drevillon}
\affiliation{Institut Pprime, Universit\'e de Poitiers-CNRS-ENSMA, UPR 3346,  ENSIP B\^atiment B25, 2 Rue Pierre Brousse, TSA 41105, 86073 Poitiers Cedex 9, France}
\author{Philippe Ben-Abdallah}
\affiliation{Laboratoire Charles Fabry, UMR 8501, Institut d'optique, CNRS, Universit\'e Paris-Sud 11, 2 avenue Augustin Fresnel, 91127 Palaiseau, France}
\date{\today}
\begin{abstract}
We show in this article that phase change materials (PCM) exhibiting a phase transition between a dielectric state and a metallic state are good candidates to perform modulation as well as amplification of radiative thermal flux. We propose a simple situation in plane parallel geometry where a so-called radiative thermal transistor could be achieved. In this configuration, we put a PCM between two blackbodies at different temperatures. We show that the transistor effect can be achieved easily when this material has its critical temperature between the two blackbody temperatures. We also see, that the more the material is reflective in the metallic state, the more switching effect is realized whereas the more PCM transition is stiff in temperature, the more thermal amplification is high. We finally take the example of VO$_2$ that exhibits an insulator-metallic transition at 68$^0$C. We show that a demonstrator of a radiative transistor could easily be achieved in view of the heat flux levels predicted. Far-field thermal radiation experiments are proposed to back the results presented.
\end{abstract}

\maketitle


Thermal flow management has become a very important challenge due to the limited energy ressources and global warming issues. Indeed, engines or power plants produce large amounts of heat which are usually wasted and rejected in the atmosphere. This heat could be advantageously used to heat buildings or domestic water. Moreover, numerous devices components and objects need to be maintained at a well established temperature in order to work properly and efficiently. If tools already exist to manage heat such as heat pipe, heating and cooling devices, there is nothing in thermal science which allows to manage heat flows in a way electricity can be guided, amplified or modulated.  

Mastering electronic transport has been possible through the discovery of two components : the electronic diode and the transistor. Diodes are able to regulate a tension whereas the transistor has three main functions : it can act as a switch, an amplifier or a modulator. The discovery of PN junction in semiconductors has paved the way to the miniaturization of such elementary components and has been one of the biggest scientific and technological breakthrough in the 20th century. It has allowed conceiving and making of circuits that now control most aspects of everydaylife.

Thermal rectification can be defined as an asymmetry in the heat flux when the temperature difference at the end of a thermal component is reversed. Thermal rectifiers have been proposed mainly in conductors. Rectification is obtained by managing phonon or heat carrier transport making the phonon or electron flux asymmetric in the material \cite{Terraneo:2002tr,Li:2004vt,Li:2005vt,Chang:2006fr,Hu:2006wk,Hu:2009vg,Yang:2007ua,Segal:2008ws,Yang:2009wt}. It can be used to conceive and realize transistors leading to the possibility of logical thermal circuits \cite{Wang:2007uf,Lo:2008tc}. More recently, radiative thermal rectifiers have been proposed \cite{Otey:2010if,Basu:2011ur,BenAbdallah:2013jp,Nefzaoui:2014ec,Nefzaoui:2014hw,ito:2014gf} as well as a radiative thermal transistor \cite{BenAbdallah:2014jk}. Very promising devices have been proposed based on the use of Vanadium dioxyde (VO$_2$) especially in the near-field \cite{BenAbdallah:2013jp,BenAbdallah:2014jk,vanZwol:2012ti}. If the rectification ratio achieved and the flux exchanged are very promising, experiments to demonstrate transistor effects in the near-field are quite difficult since materials have to be maintained at subwavelength scale i.e at micronic distances at ambient temperature. The goal of this article is to reconsider in the far-field the configuration proposed in the near-field in order to propose experiments that would be able to demonstrate thermal transistor effects. We first recall the case of a system composed of a phase change material (PCM) and a blackbody. We derive an expression for the rectification ratio characterizing such a system. We discuss with simple considerations the properties that have to be exhibited by PCMs in order to achieve a good rectification. Then, we consider the case of three bodies in plane-parallel configuration : a PCM is introduced between two blackbodies maintained at different temperatures. We show that transistor effect can be achieved if the material critical temperature is close to the equilibrium temperature of a radiative screen. We show that a good switch is realized when the ratio between the emissivity in the metallic and dielectric phase is very low. Concerning the amplification, the condition is different since amplification will be high if the temperature range on which transition occurs is small. We finally take the example of VO$_2$ and show that transistor effect should be observed with this material in simple far-field experiments.



We first consider a system where a plane-parallel material at temperature $T_1$ is exchanging heat by thermal radiation with a second material at temperature $T_2$. Both materials are opaque. This configuration has already been introduced by Ben-Abdallah and Biehs in 2013 \cite{BenAbdallah:2013jp}. If the distances involved are much larger than the thermal wavelength, heat transfer is limited to the classical propagative waves and no tunneling occurs : heat flux density exchanged between the two materials reduces to \cite{Mulet:2002we,Joulain:2005ih}
\begin{eqnarray}
\label{flux12}
\phi_{12}(T_1,T_2)=\int_{\Omega=2\pi}\!\!\!\cos\theta  d\Omega\int_0^\infty d\lambda\left[L^0(T_1,\lambda)-L^0(T_2,\lambda)\right]\nonumber\\
\times\frac{1}{2}\sum_{i=s,p}\frac{(1-|r_1^i(T_1,\lambda,\theta)|^2)(1-|r_2^i(T_2,\lambda,\theta)|^2)}{|1-r_1^i(T_1,\lambda,\theta)r_2^i(T_2,\lambda,\theta)\exp{2i k_0 \sin\theta d}|^2}
\end{eqnarray}
where $L^0(T,\lambda)$ is the specific intensity of a blackbody at temperature $T$ and for a wavelength $\lambda$, $r_j^i$ is the Fresnel reflexion coefficient between vacuum and material $j$ with polarization $i$ ($s$ or $p$), $d$ is the separation distance between the two interfaces and $k_0=2\pi c/\lambda$ is the wave vector in vacuum, $c$ being the light velocity. Integration is also taken on the solid angle $d\Omega=2\pi\sin\theta d\theta$ and $\theta$ is the angle with the perpendicular direction to the two media. The fact that $d\gg\lambda$ make the interference term in the denominator vanish \cite{Mulet:2002we}. It is in that case possible to introduce an emissivity $\varepsilon_j^i=1-|r_j^i|^2$ and a reflectivity $|r_j^i|^2$. In case of a lambertian source i.e. classical thermal sources where the emissivity does not depend on the angle and the polarization, integration over $\Omega$ is $\pi$. (\ref{flux12}) reduces to 
\begin{eqnarray}
\label{flux12_lamb}
\phi_{12}(T_1,T_2)=\int_0^\infty\frac{\varepsilon_1(T_1,\lambda)\varepsilon_2(T_2,\lambda)}{1-\rho_1(T_1,\lambda)\rho_2(T_1,\lambda)}\nonumber\\\times\left[H^0(T_1,\lambda)-H^0(T_2,\lambda)\right]d\lambda
\end{eqnarray}
where $H^0_\lambda(T)=\pi L^0_\lambda$ is the emittance.
Rectification is observed when $\phi_{12}(T_1,T_2)$ is different in modulus from $\phi_{12}(T_2,T_1)$. Examination of Eq.(\ref{flux12}) and  (\ref{flux12_lamb}) shows that rectification is observed if and only if materials are different and exhibit a variation of their optical properties with temperature which is different for the two materials. Rectification is defined by the parameter $R$

\begin{equation}
\label{ R_Ratio}
R(T_1,T_2)=\frac{|\phi_{12}(T_1,T_2)+\phi_{12}(T_2,T_1)|}{Max(|\phi_{12}(T_1,T_2)|,|\phi_{12}(T_2,T_1)|)}
\end{equation}
The simplest and more efficient way to achieve thermal rectification is to consider a blackbody material which emissivity is equal to 1 over the whole spectrum whatever is the temperature and a PCM exhibiting a rapid phase transition at a temperature $T_c$. Both bodies are considered as lambertian. 

Let us therefore consider that material 1 is a blackbody material and that material 2 is a grey PCM with an emissivity $\varepsilon$ equal to $\varepsilon^{<}$ for $T\leq T_c$ whereas it is equal to $\varepsilon^{>}$ for $T>T_c$. The forward flux can easily be written as $\phi_{12}(T_1,T_2)=\varepsilon(T_2)\sigma(T_1^4-T_2^4)$ whereas the backward flux is equal to  $\phi_{12}(T_2,T_1)=\varepsilon(T_1)\sigma(T_2^4-T_1^4)$. Rectification follows as
\begin{equation}
\label{ }
R(T_1,T_2)=1-\frac{\min(\varepsilon(T_1),\varepsilon(T_2))}{\max(\varepsilon(T_1),\varepsilon(T_2))}
\end{equation}
In this simple situation, one note once again that rectification is achieved as long as the emissivity depends on the temperature. The more the emissivity contrast is large, the more rectification will be important. In the example where $T_1$ is set to a constant value and the critical temperature $T_c>T_1$, then $R(T_1,T)$ is equal to 0 if $T<T_c$ and to $1-\min(\varepsilon^{<},\varepsilon^{>})/\max(\varepsilon^{<},\varepsilon^{>})$ when $T>T_c$. From these considerations, one can infer that materials exhibiting a transition from a state where it is highly reflective (a metal where $\varepsilon^>=\varepsilon^{met}$) to a state where it is a strong absorbing material (a dielectric where $\varepsilon^<=\varepsilon^{diel}$) will be excellent to exhibit radiative thermal rectification. Rectification ratio then read $R=1-\varepsilon^{met}/\varepsilon^{diel}$. 

The ideal case would be a material that behaves as a perfect reflector in one phase, this kind of material being able to stop radiative flux, whereas it would be a blackbody in a another phase at a different temperature. This kind of material would make a perfect thermal diode. The deviation from the ideal case can actually be measured as the ratio $\varepsilon^{met}/\varepsilon^{diel}$. Thus, thermochromic materials and Mott insulator metallic transition materials such as VO$_2$ are good candidates to perform thermal rectification \cite{BenAbdallah:2013jp}.

\begin{figure}
\begin{center}
\includegraphics[width=8cm]{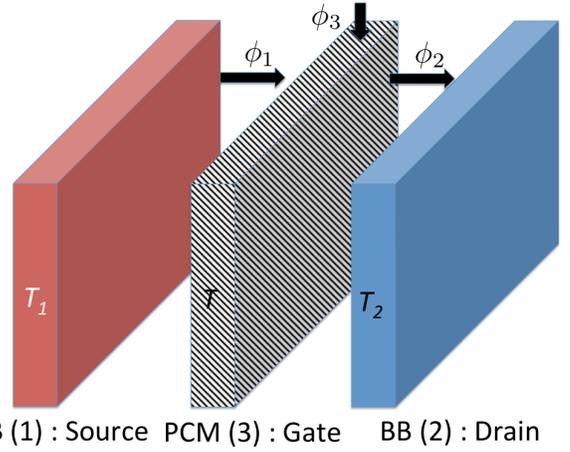}
\caption{Transistor geometry : a phase change material (PCM) exhibiting a transition phase at $T_c$ is placed between two blackbodies (BB) at $T_1$ and $T_2$.}
\label{Transist_geom}
\end{center}
\end{figure}

Such a transition can be exploited in order to make a transistor as it has been suggested by Ben-Abdallah and Biehs \cite{BenAbdallah:2014jk} in near-field situations. Following this idea, it is possible to propose such a radiative thermal transistor in far-field plane-parallel geometry provided that the material exhibits a transition for its emissivity (Fig. \ref{Transist_geom}). 

Let us put a planar PCM which is maintained at temperature $T$ by an external thermal source between two bodies at temperatures $T_1$ and $T_2$ with $T_1>T_2$. Let us call $\phi_1$ the heat flux density exchanged between material 1 and the PCM at temperature $T$ and $\phi_2$ the heat flux density exchanged between the PCM and material 2. $\phi_3$ is the heat flux received by the material in order to be maintained at temperature $T$. Thus, expressions of these three fluxes can be written with expression (\ref{flux12})
\begin{eqnarray}
\phi_1 & = & \phi_{13}(T_1,T) \nonumber\\
\phi_2 & = &  \phi_{31}(T,T_2)\\
\phi_3 & = & \phi_2-\phi_1 \nonumber 
\end{eqnarray}

\begin{figure}
\begin{center}
\includegraphics[width=8cm]{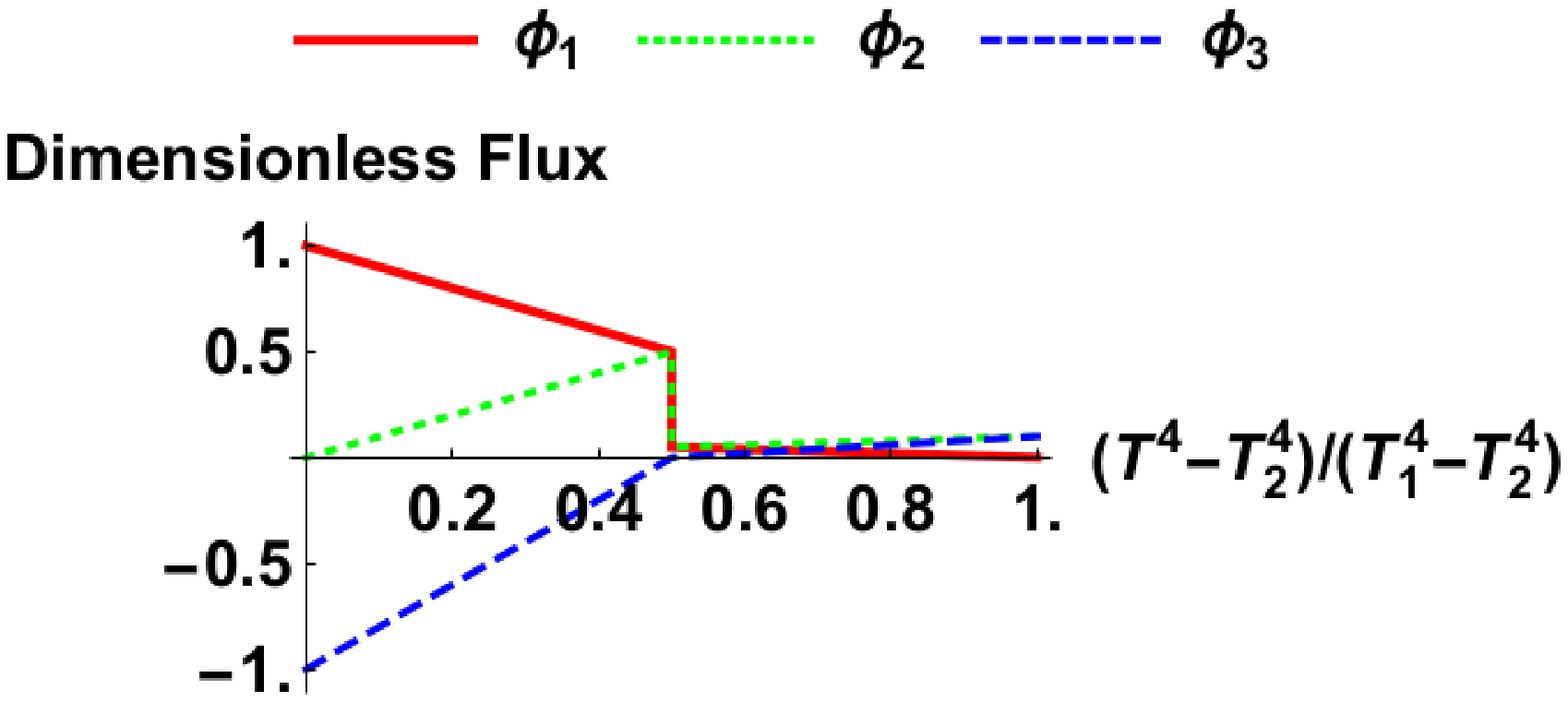}
\includegraphics[width=8cm]{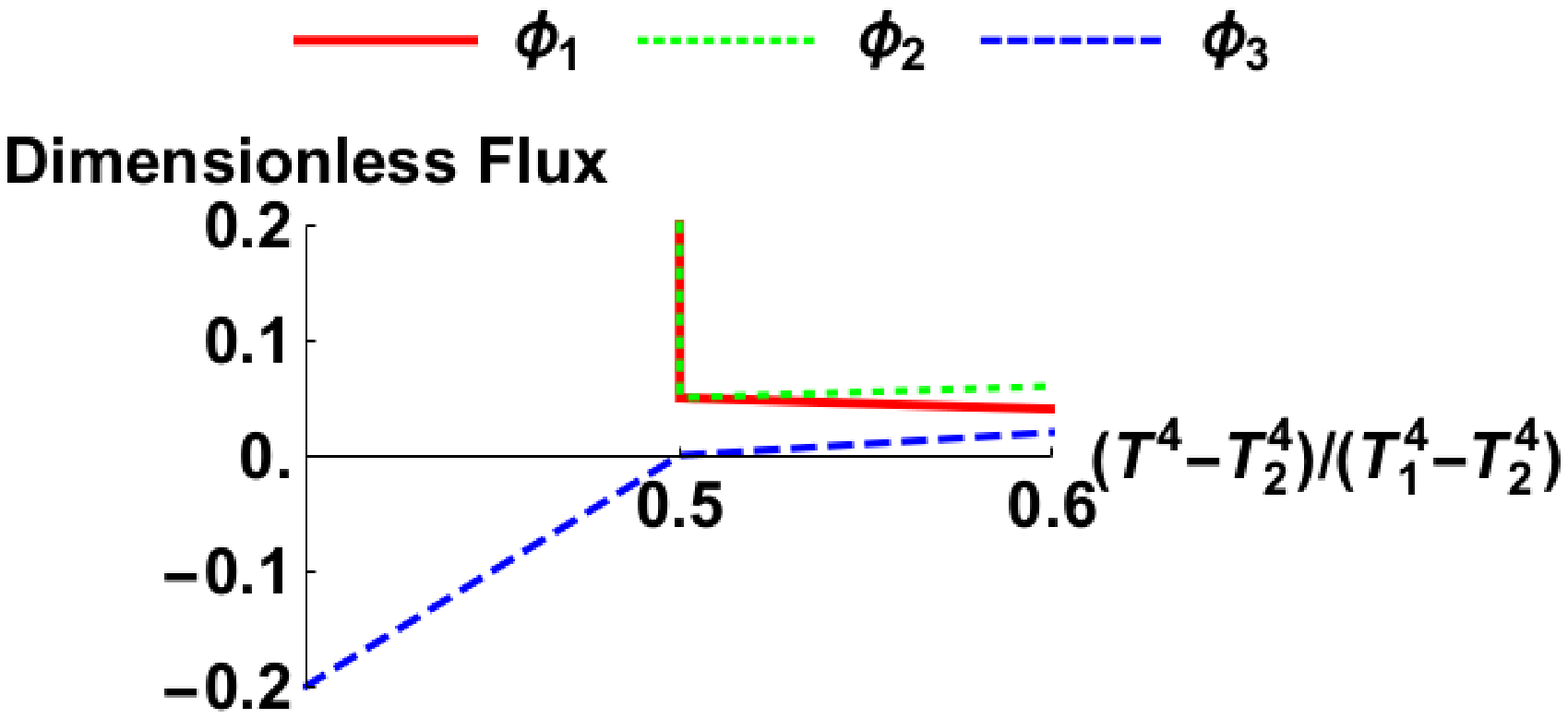}
\caption{Upper figure : Dimensionless fluxes in the thermal transistor vs $T^4-T_2^4/(T_1^4-T_2^4)$. $T_c$ is such that $T_c=T_{eq}=[0.5(T_1^4+T_2^4)]^{1/4}$. Here $\varepsilon^<=\varepsilon^{diel}=1$ and $\varepsilon^>=\varepsilon^{met}=0.1$. A slight increase of the external flux $\phi_3$ received by the phase change material (Gate) provokes a strong decrease of the fluxes $\phi_1$ and $\phi_2$ (Source and Drain thermal currents) whereas a slight decrease leads to a strong increase of the fluxes. Lower figure : zoom of the above figure in the transition zone.}
\label{adim}
\end{center}
\end{figure}

All the behavior of such a system can be understood in the simple case of the phase change material considered above put between two blackbodies. Preceding equations on the heat flux densities now read
\begin{eqnarray}
\phi_1 & = & \varepsilon(T)\sigma(T_1^4-T^4) \\
\phi_2 & = & \varepsilon(T)\sigma(T^4-T_2^4) \\
\phi_3 & = &\phi_2-\phi_1=\varepsilon(T)\sigma(2T^4-T_1^4-T_2^4)
\end{eqnarray}
As an exemple, let us take a material which is a blackbody below $T_c=[0.5(T_1^4+T_2^4)]^{1/4}$ ($\varepsilon^<=\varepsilon^{diel}=1$) and has an emissivity equal to 0.1 above $T_c$ ($\varepsilon^>=\varepsilon^{met}=0.1$). In Fig. \ref{adim}, we report the three fluxes defined above that have been normalized to the quantity $\sigma(T_1^4-T_2^4)$ as functions of $(T^4-T_2^4)/(T_1^4-T_2^4)$. We note that at thermal equilibrium $T^4=0.5(T_1^4+T_2^4)=T_{eq}^4$ and $\phi_3=0$. When the flux $\phi_3$ is slightly increased in order that $T$ exceeds the transition temperature $T_c$, thermal fluxes $\phi_1$ and $\phi_2$ are drastically reduced. On the contrary, a slight decrease in the heat flux $\phi_3$ provokes a important increase of thermal fluxes $\phi_1$ and $\phi_2$ when $T$ goes below $T_c$. 

Following \cite{BenAbdallah:2014jk}, this configuration can therefore be seen as a thermal analog of a field effect transistor when blackbody 1 would play the role of the source, blackbody 2 the role of the drain whereas the PCM plays the role of the gate. This thermal transistor can therefore control a thermal radiative heat flux between the source and the drain acting on the thermal flux $\phi_3$ that control the gate temperature. Depending on the thermal properties variation of the PCM, this device can be an efficient thermal modulator or even a thermal switch if the metallic phase is sufficiently reflective. Note that the circuit is switched off only if the material is perfectly reflecting. Note also that this transistor effect is better achieved when the critical temperature $T_c$ is close to the equilibrium temperature $T_{eq}$ as it has been chosen here. In this case, one only needs a small amount of heat to switch the PCM from one state to another. When it is not the case, transition occur for a larger amplitude of the amount of heat brought to the PCM.

Note that this device actually exhibits a negative differential thermal conductance which is characteristic of a transistor. Such a behavior is observed when the quantity defined as
\begin{equation}
\label{ }
\alpha=\left|\frac{\partial \phi_2}{\partial\phi_3}\right|=\frac{1}{|1-(d\phi_2/dT)/(d\phi_3/dT)|}
\end{equation}
is greater than 1. Indeed, $\alpha$ greater than 1 means that an increase or decrease of $\phi_3$ (excitation) leads to an increase or decrease of $\phi_2$  (response) greater than the excitation which is the definition of an amplification.
In the current modeling, $\alpha$ reads
\begin{equation}
\label{ }
\alpha=\frac{1}{|1-\frac{\varepsilon'(T_1^4-T^4)-4\varepsilon T^3}{\varepsilon'(T^4-T_2^4)+4\varepsilon T^3}|}
\end{equation}
where $\varepsilon'=d\varepsilon/dT$. Let us note that outside the transition zone, $\varepsilon'=0$ so that $\alpha=1/2$. In the transition zone, emissivity typically varies of $\Delta\varepsilon$ on a thermal interval $\Delta T$. If this transition is sufficiently steep so that $\Delta\varepsilon/\varepsilon\gg\Delta T/(T_1-T)$ then $\alpha\simeq1/|1-(T_1^4-T_c^4)/(T_c^4-T_2^4)|$. Note that $\alpha$ is greater than 1 in the transition zone which leads to an amplification of thermal currents $\phi_1$ and $\phi_2$ by $\phi_3$ around $T_c$. Note also that if $T_c=T_{eq}$, the preceding expression of $\alpha$ diverges and is not correct. Asymptotic development around $T_c$ gives 
\begin{equation}
\label{ }
\alpha\simeq\frac{\Delta\varepsilon(T_1-T_c)}{\Delta T(\varepsilon^{met}+\varepsilon^{diel})}
\end{equation}
Contrary to the switch effect which needs an emissivity as small as possible, amplification needs a strong emissivity variation on a small temperature range. Thus, it is the stiffness of the transition $\Delta\varepsilon/\Delta T$ which will determine how strong amplification $\alpha$ will be. As amplification exists only if $\alpha$ is larger than 1, it will occur only if $\Delta\varepsilon/\Delta T$ is larger than $(\varepsilon^{met}+\varepsilon^{diel})/(T_1-T_c)$. Therefore, a weak stiffness in the transition can be compensated by a large temperature difference between the source and the gate.


Vanadium dioxyde (VO$_2$) has been the subject of numerous researches due to its remarkable properties and particularly its insulator-metal transition around 68$^0$C \cite{Morin:1959uc,Biermann:2005ts,Qazilbash:2007cm,Qazilbash:2009wb}. This material is an excellent candidate to actually achieve both thermal rectification \cite{BenAbdallah:2013jp,ito:2014gf} and a radiative thermal transistor in the far field. VO$_2$ is quite easy to grow and to be deposited on a substrate. Its temperature can be easily controlled  by optical or electrical means \cite{Qazilbash:2009wb} as fast as 100 fs \cite{Rini:2005ts}. Its transition temperature can also be shifted to room temperature by doping it with W \cite{Paradis:2007tx}.  

\begin{figure}
\begin{center}
\includegraphics[width=8cm]{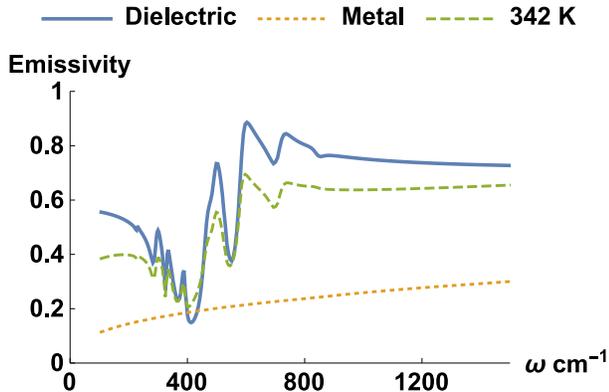}
\caption{VO$_2$ emissivity spectral dependance with wavenumber $\omega$ (in cm$^{-1}$) in the dielectric phase ($T<340$ K), in the metallic phase $T>345$ K and in the phase transition temperature range $T=342$ K using the Bruggeman model.}
\label{emisvo2}
\end{center}
\end{figure}

In the plane-parallel geometry presented in the preceding section, we have calculated performances that would be, if realized, among the best transistor ever constructed. To perform our calculations, we have used actual optical properties of VO$_2$. In the phase transition temperature zone [341 K, 345 K], we have taken the Bruggeman mixing rule that have been introduced by Qazilbash et al \cite{Qazilbash:2007cm,Qazilbash:2009wb}. VO$_2$ emissivity spectral variations for a plane interface are represented in Fig. \ref{emisvo2}. We see that emissivity is greatly reduced when the material temperature is above critical temperature $T_c$ so that it is in the metallic phase. In the temperature range where the transition occurs, the material emissivity remains between the one of the two pure phases.

\begin{figure}
\begin{center}
\includegraphics[width=8cm]{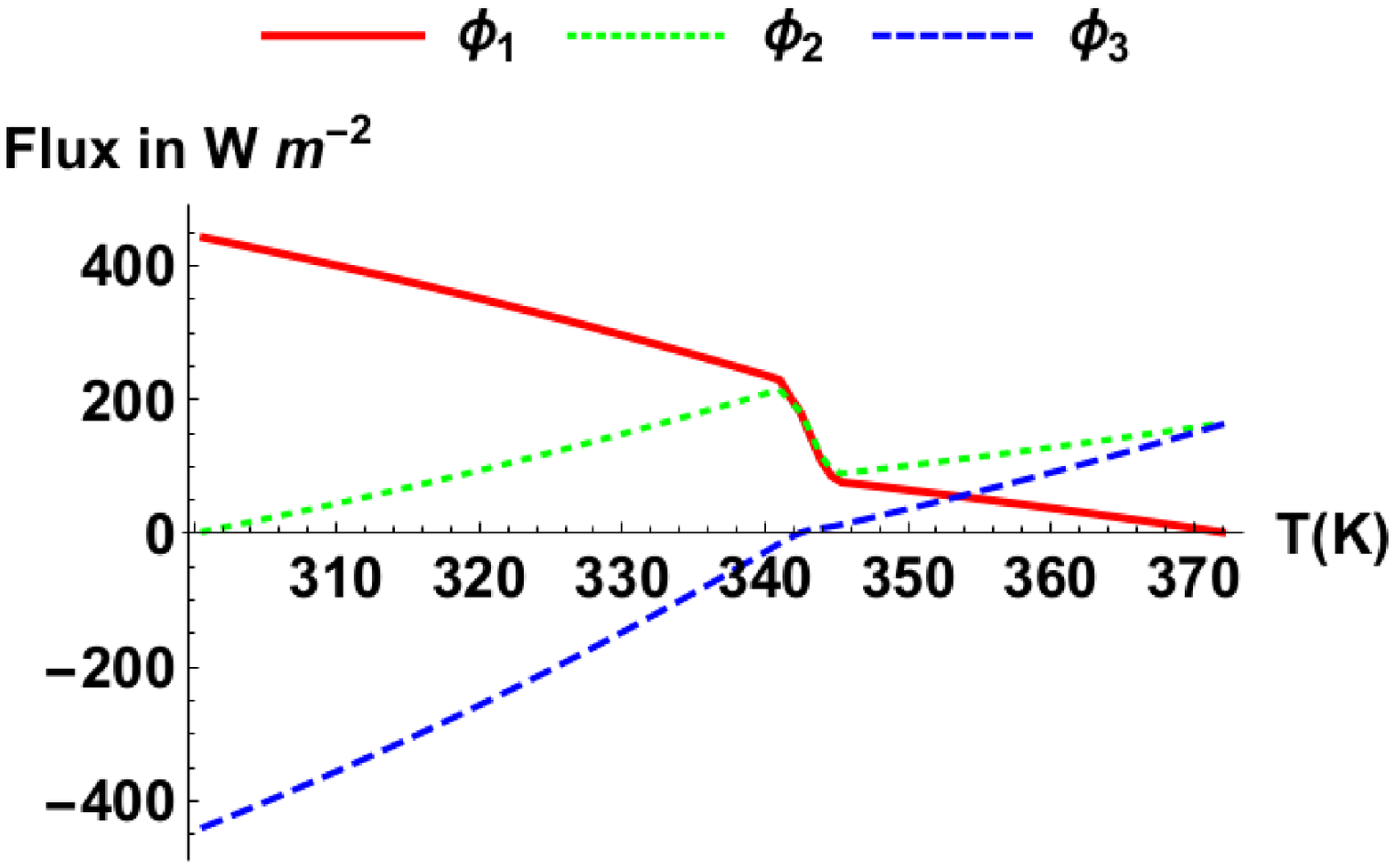}
\includegraphics[width=8cm]{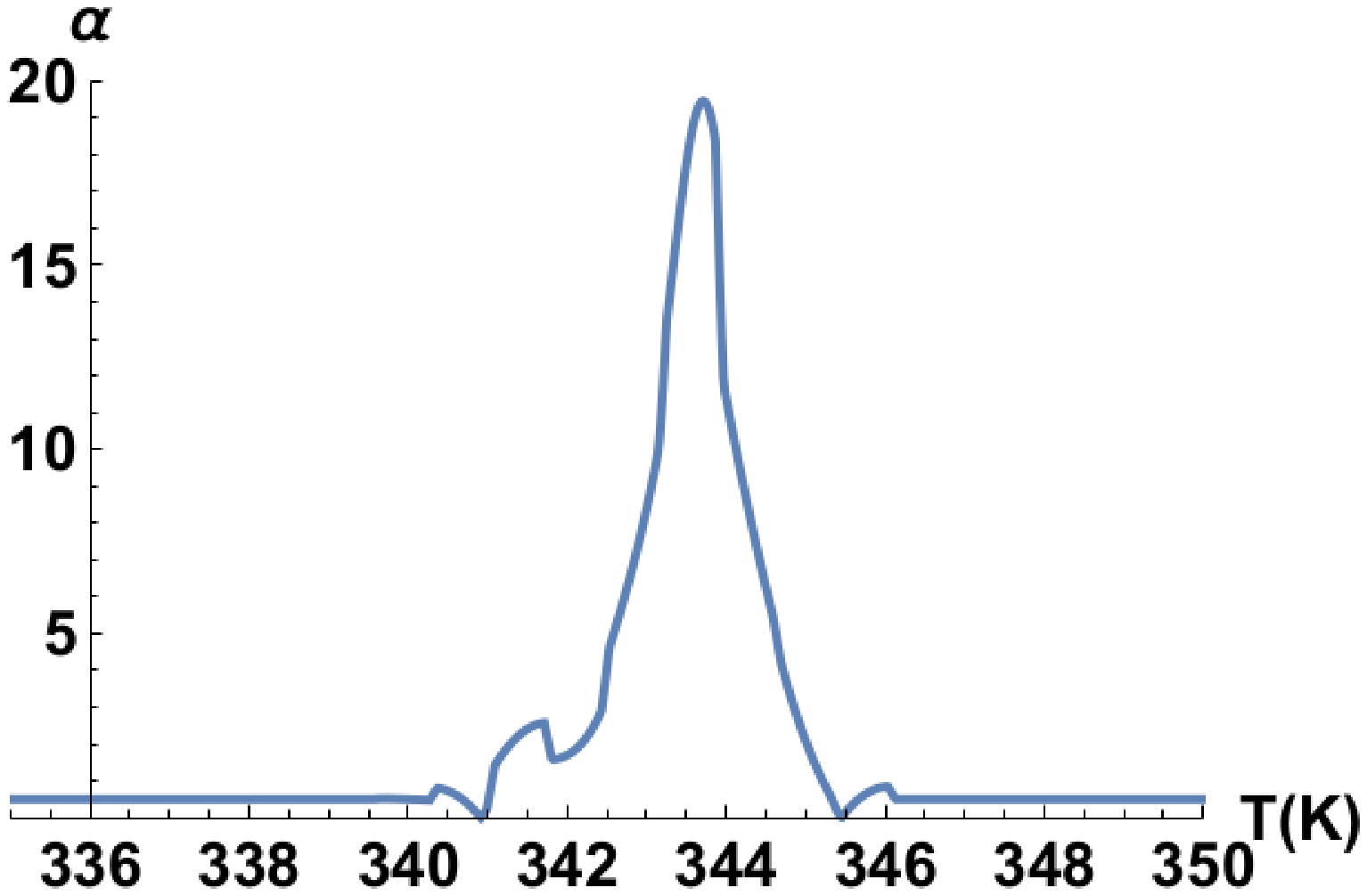}
\caption{Upper figure : Radiative flux in a radiative thermal transistor made of two blackbodies and VO$_2$. The two blackbodies are supposed to be at temperature $T_1$= 372.5 K and $T_2$= 300 K and the VO$_2$ is at temperature $T$. $\phi_1$ is the radiative flux between the blackbody (1) (source) and VO$_2$ (gate), $\phi_2$ is the radiative flux between the VO$_2$ (gate and blackbody (2) (drain). $\phi_3=\phi_2-\phi_1$ is the flux required to maintain VO$_2$ at temperature $T$. Lower figure : amplification factor $\alpha$ of the radiative heat transfer. The amplification factor is above 1 in the phase transition temperature range between 341 K and 345 K.}
\label{transvo2}
\end{center}
\end{figure}

We consider the case where $T_1$ =  372.5 K and $T_2$ = 300 K. When the PCM is not temperature controlled ($\phi_3=0$), it reaches an equilibrium temperature $T_{eq}=[(T_1^4+T_2^4)/2]^{1/4}$= 342 K that lies into the phase transition temperature range. In Fig. \ref{transvo2}, we reported the three fluxes versus the PCM (gate) temperature in a similar way it has been plotted in Fig. \ref{adim} for an idealized transistor with a sharp phase transition. We indeed see that $\phi_3=0$ at $T=342$ K. We also note that a slight increase in thermal flux $\phi_3$ leads to a sharp decrease of the source-gate and gate-drain radiative flux. This is the signature of the transistor effect which is also confirmed when one calculates the radiative flux amplification factor $\alpha$. We note in this figure that the amplification factor is above 1 in the temperature range where the phase transition occurs between 341 K and 345 K and reaches values slightly lower than 20. 

An interesting point is also to look at the level at which radiative heat transfer is exchanged between the different bodies. Typical radiative heat flux densities are between a few tens and a few hundreds of W m$^{-2}$. This levels are 2 orders of magnitude lower than the one obtained in the near-field  \cite{BenAbdallah:2014jk}. However, experiments in the far-field are much easier to perform since the separation distances between bodies have not to be managed at a submicronic scale. Thermal radiation density flux levels predicted in far-field are frequently measured in all natural convection experiments or in typical thermal building science experiments. On the other way, natural convection will of course perturbate measurements in such a device. Typical natural convection fluxes being of the same order of magnitude than radiative flux, this kind of experiment will preferentially be operated in vacuum. This kind of experiment has been recently done to prove the rectification effect with a VO$_2$ sample \cite{ito:2014gf}. In similar conditions, far field-radiation experiments between two blackbodies and a sample of VO$_2$ in vacuum should allow proof of this thermal transistor effect.

We have shown in this article that phase change materials put between two blackbodies in plane-parallel geometry are good candidates to fabricate a radiative thermal transistor in the far-field able to modulate, switch or amplify radiative heat flux. Switching performances are driven by the ratio of the phase change material emissivity in the metallic phase and in the dielectric phase. Amplification performances mainly depends on the stiffness of the emissivity variations with the temperature. 
If the performances in the far-field are lower in terms of levels of heat exchange, this configuration appears to be much easier than in the near-field due to the difficulty and constraints of position and temperature control in this last configuration. The existence of phase change materials such as thermochromic material or Mott transition material at ambient temperature offers good opportunities to prove the possibility of making radiative heat flux transistor as shows the example of VO$_2$. This could open the way to future thermal logic gates conception and fabrication. With such thermal logical elements, it could be possible to perform a thermal control without electronic control. This could lead to a passive thermal control of elements that would be useful in situations where reliability is searched. 

\acknowledgements{This work pertains to the French Government Program ''Investissement d'Avenir'' (LABEX INTERACTIFS, ANR-11-LABX-0017-01}



%

\end{document}